\documentstyle[12pt,psfig]{article}

\begin{document} 
\title{The role of the $P_{11}(1710)$ in the 
$NN{\to}N{\Sigma}K$ reaction\thanks{Supported in part by the
Forschungszentrum J\"ulich and the Australian Research Council}}
\author{
A. Sibirtsev$^1$~\thanks{alexandre.sibirtsev@theo.physik.uni-giessen.de} ,
K. Tsushima$^2$~\thanks{ktsushim@physics.adelaide.edu.au} ,
W. Cassing$^1$~\thanks{wolfgang.cassing@theo.physik.uni-giessen.de} ,
A. W. Thomas$^2$~\thanks{athomas@physics.adelaide.edu.au} \\
{ $^1$\small Institut f\"ur Theoretische Physik, Universit\"at Giessen} \\
{\small D-35392 Giessen, Germany} \\
{\small $^2$Center for the Subatomic Structure of Matter (CSSM)} \\
{\small and Department of Physics and Mathematical Physics } \\
{\small University of Adelaide, SA 5005, Australia}}
\date{}
\maketitle

\vspace{-8.0cm}
\hfill UGI-98-36,\quad ADP-98-68/T335
\vspace{8.0cm}

\begin{abstract}
Using the resonance model, which was successfully applied for 
the study of the $pp{\to}p{\Lambda}K^+$ reaction, we investigate  
$NN{\to}N{\Sigma}K$ reactions that are expected
provide cleaner information about resonance 
excitations and meson exchange contributions. For this purpose we 
demonstrate 
that the invariant mass distribution for the ${\Sigma}K$ system, as 
well as the Dalitz plot for the $NN{\to}N{\Sigma}K$ reaction,   
provide direct information about the ${\Sigma}K$ production 
mechanism, which can be tested in the near future by experiments at COSY.
\end{abstract}

\vspace{2cm}
PACS: 13.30.-a; 13.75.Ev, 13.75.Cs

\vspace{1cm}
Keywords: Decays of baryons; Nucleon-nucleon interactions;
Strangeness production; Hyperon-nucleon interactions.

\newpage
\section{Introduction}
The mechanism of strangeness production in nucleon-nucleon ($NN$)
reactions is still an open problem. In 1960 Ferarri~\cite{Ferrari}  
was the first to propose kaon and pion exchanges as a 
mechanism for strangeness production in $NN$ reactions. 
It is obvious that one can extend this approach by incorporating other   
nonstrange mesons, $\pi$, $\eta$, $\sigma$, $\rho$ and $\omega$ as 
well as the known strange mesons, $K$, $K^\ast$ and $K_1$.
In particular, the $\pi$, $\eta$ and $\rho$ mesons are expected to 
give large contributions because recent experimental data show 
that they couple strongly to several baryonic  resonances ($R$) 
which have sizeable decay branches to a hyperon ($Y$) and a kaon ($K$).
This is especially apparent in the case of $\pi N \to Y K$ 
reactions~\cite{Tsushima1}. Furthermore, among the different meson 
exchanges the $K$-meson exchange is of fundamental interest
because the kaon-nucleon-hyperon  coupling constant, $g_{KNY}$, 
is related to the strength for $s\bar{s}$-pair creation at the 
$KNY$ vertex.  
However, as was found in several studies~\cite{Laget,Deloff,
Li,Sibirtsev1}, it is not easy to separate the contributions 
from strange and nonstrange meson exchanges on the basis of the
available data. It was shown in Refs.~\cite{Yao,Wu,Tsushima2} that 
to reproduce the available experimental data the 
kaon exchange  may not be necessary.

In Ref. \cite{Sibirtsev2} we proposed a possibility  
to distinguish the dominant meson exchanges in 
the strangeness production, $NN{\to}N{\Lambda}K$, experimentally 
by reconstructing the invariant mass distribution for
the ${\Lambda}K$ system. It was found that the strangeness production 
in the $NN{\to}NYK$ reaction could be understood in terms of  
nonstrange meson exchanges followed by the excitation of an 
intermediate baryon resonance, $R$, which  
decays to a $\Lambda$ and kaon. 

Thus, the $YK$ invariant mass spectrum is a promising 
way to understand the structure of strangeness production 
especially with respect to baryonic resonance excitation.
Furthermore, we note that at low energies strange meson exchange
provides almost the same invariant mass spectrum as an 
isotropic phase-space distribution for  
the hyperon-kaon system in the final state. Therefore, there is the 
possibility to distinguish the kaon exchange contributions from 
$\pi$, $\eta$ and $\rho$-meson exchanges, which 
deviate from isotropic distributions, if they lead to strong
baryonic resonance excitations.

Recently, the TOF Collaboration from the COoler SYnchrotron, 
COSY-J\"ulich, reported on experimental data for  
the ${\Lambda}K$ invariant mass spectrum~\cite{TOF}.
They observe deviations from an isotropic phase space distribution 
and are close to the predicted spectrum~\cite{Sibirtsev2}. Obviously,
to obtain further insight into the strangeness production mechanism, 
it is crucial to collect more statistics and also to achieve a 
better mass  resolution for the ${\Lambda}K$ system. On the other 
hand, the observation 
in Ref.~\cite{TOF} can be considered as a first experimental 
indication for the excitation of an
intermediate baryonic resonance coupled to the ${\Lambda}K$
system in the $pp{\to}p{\Lambda}K^+$ reaction.

In this study we calculate the $NN{\to}N{\Sigma}K$ reaction within the
same one-meson exchange model used in 
Refs.~\cite{Tsushima1,Tsushima2,Sibirtsev2},
which includes $\pi$, $\eta$ and $\rho$-meson exchanges followed 
by intermediate state resonance excitations.
As is commonly believed, the kaon exchange contribution to $\Sigma$ 
production in the $NN$ reaction  is almost negligible due to the 
small $KN{\Sigma}$ coupling constant (cf. Ref.~\cite{Sibirtsev3}).
Thus we expect the signal due to resonance excitations 
in the $\Sigma$ channel in $NN$ reactions, $R{\to}{\Sigma}K$, 
to be much cleaner than in the $\Lambda$ production channel.  
Furthermore, in  case of the $pp{\to}p{\Lambda}K^+$ reaction 
there are a few resonances which  
contribute and  therefore smear out
the invariant mass distribution for the ${\Lambda}K$ system. 
On the other hand, as will be shown later, the situation for 
$\Sigma$ production is much better because the dominant 
contribution mainly should stem from the $N(1710)$ resonance.

\section{The model}

\begin{figure}[ht]
\psfig{file=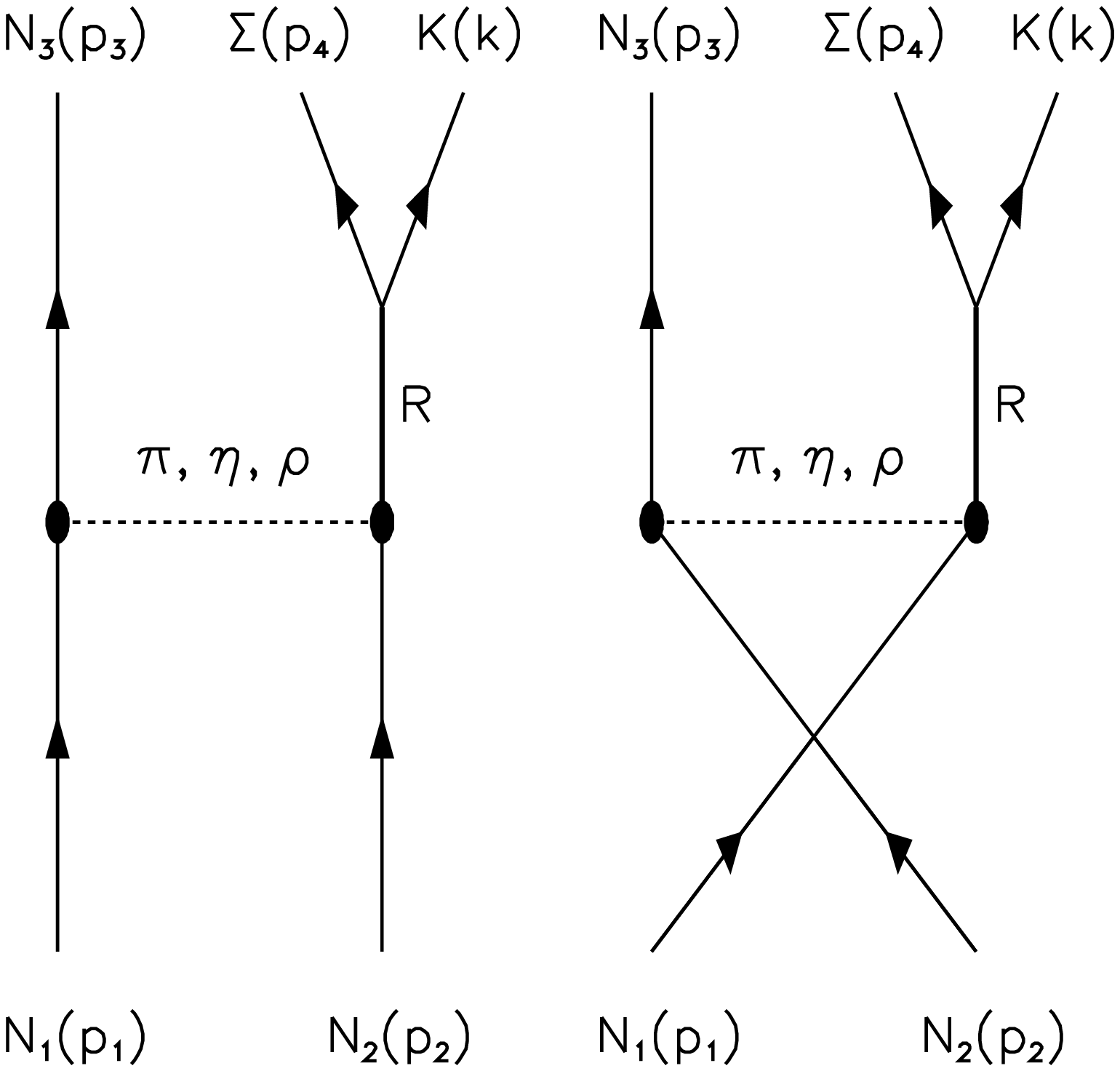,width=16cm}
\caption[]{\label{kn0}Feynman diagrams for the reaction
$NN{\to}N{\Sigma}K$ considered in the model.}
\end{figure}

The Feynman diagrams relevant for the $NN{\to}N{\Sigma}K$
reaction are shown in Fig.~\ref{kn0}. Here $R$ denotes a
baryonic resonance which couples to the ${\Sigma}K$ channel,
while $M$ denotes the exchanged meson ($\pi$, $\eta$ 
or $\rho$). The mesons are restricted to those which are 
observed in the decay channels of the resonances considered.

Since we only take into account those resonances which 
are experimentally known to decay to a $\Sigma$ and kaon, 
and also the branching ratios to the $\pi N$, $\eta N$ and $\rho N$ modes 
are experimentally measured, 
all the relevant coupling constants at the $MNR$ and $K \Sigma R$
vertices can be determined from the measured partial widths.
The data for the resonances included in the model
are summarized in Table~\ref{tab2}. 
Note that some of the data have been modified in Ref.~\cite{PDG1}.  
The $\Delta(1920)$ resonance 
in the model is treated as an effective resonance which represents the 
contributions from six individual resonances, 
i.e. $\Delta(1900)$, $\Delta(1905)$, $\Delta(1910)$,
$\Delta(1920)$, $\Delta(1930)$ and $\Delta(1940)$ in line with
Ref.~\cite{Tsushima1}. 

\begin{table}
\caption{\label{tab2}
The properties of the baryonic resonances included in the model.
Confidence levels of the resonances are, $N(1710)^{***}$, 
$N(1720)^{****}$ and $\Delta(1920)^{***}$~\protect\cite{PDG}.}
\begin{center}
\begin{tabular}{|c|cccc|}
\hline
Resonance $(J^P)$ &Width (MeV) &Decay channel &Branching ratio 
&Adopted value \\
\hline
N(1710)$\,(\frac{1}{2}^+)$ &100 &$N \pi$      &0.10 -- 0.20 &0.150 \\
                           &    &$N \eta$     &0.20 -- 0.40 &0.300 \\
                           &    &$N \rho$     &0.05 -- 0.25 &0.150 \\
                           &    &$\Sigma K$   &0.02 -- 0.10 &0.060 \\
\hline
N(1720)$\,(\frac{3}{2}^+)$ &150 &$N \pi$      &0.10 -- 0.20 &0.150 \\
                           &    &$N \eta$     &0.02 -- 0.06 &0.040 \\
                           &    &$N \rho$     &0.70 -- 0.85 &0.775 \\
                           &    &$\Sigma K$   &0.02 -- 0.05 &0.035 \\
\hline
$\Delta$(1920)$\,(\frac{3}{2}^+)$ &200 &$N \pi$ &0.05 -- 0.20 &0.125 \\
                           &    &$\Sigma K$   &0.01 -- 0.03 &0.020 \\
\hline
\end{tabular}
\end{center}
\end{table}

The effective Lagrangian densities relevant for the present  
study are taken as ~\cite{Tsushima1,Tsushima2,Sibirtsev2} 
\begin{eqnarray}
{\cal L}_{\pi N N} &=& 
-ig_{\pi N N} \bar{N} \gamma_5 \vec\tau N \cdot \vec\pi,
\label{pnn}\\
{\cal L}_{\eta N N} &=& 
-ig_{\eta N N} \bar{N} \gamma_5 N \eta,
\label{enn}\\
{\cal L}_{\rho N N} &=& - g_{\rho N N}
\left( \bar{N} \gamma^\mu \vec\tau N \cdot \vec\rho_\mu
+ \frac{\kappa}{2 m_N} \bar{N} \sigma^{\mu \nu}
\vec\tau N \cdot \partial_\mu \vec\rho_\nu \right),
\label{rnn}
\end{eqnarray}
\begin{eqnarray}
{\cal L}_{\pi N N(1710)} &=&  
-ig_{\pi N N(1710)}
\left( \bar{N}(1710) \gamma_5 \vec\tau N \cdot \vec\pi
+ \bar{N} \gamma_5 \vec\tau N(1710) \cdot \vec\pi\,\, \right), 
\label{pnb}\\
{\cal L}_{\eta N N(1710)} &=&
-ig_{\eta N N(1710)}
\left( \bar{N}(1710) \gamma_5 N \eta
+ \bar{N} \gamma_5 N(1710) \eta\,\, \right),
\label{enb}\\
{\cal L}_{\rho N N(1710)} &=&
-g_{\rho N N(1710)}
\left( \bar{N}(1710) \gamma^\mu \vec\tau N \cdot \vec\rho_\mu
+ \bar{N} \gamma^\mu \vec\tau N(1710) \cdot \vec\rho_\mu\,\, \right),
\label{rnb}\\
{\cal L}_{\pi N N(1720)} &=&
\frac{g_{\pi N N(1720)}}{m_\pi}
\left( \bar{N}^\mu(1720) \vec\tau N \cdot \partial_\mu \vec\pi
+ \bar{N} \vec\tau N^\mu(1720) \cdot \partial_\mu \vec\pi \, \right),
\label{pnc}\\
{\cal L}_{\eta N N(1720)} &=&
\frac{g_{\eta N N(1720)}}{m_\eta}
\left( \bar{N}^\mu(1720) N \partial_\mu \eta
+ \bar{N} N^\mu(1720) \partial_\mu \eta \, \right),
\label{enc}\\
{\cal L}_{\rho N N(1720)} &=&
-ig_{\rho N N(1720)}
\left( \bar{N}^\mu(1720) \gamma_5 \vec\tau N \cdot \vec\rho_\mu
+ \bar{N} \gamma_5 \vec\tau N^\mu(1720) \cdot \vec\rho_\mu\,\, \right),
\label{rnc}\\
{\cal L}_{\pi N \Delta(1920)} &=& 
\frac{g_{\pi N \Delta(1920)}}{m_\pi}
\left( \bar{\Delta}^\mu(1920) \overrightarrow{\cal I} N \cdot 
\partial_\mu \vec\pi + \bar{N} {\overrightarrow{\cal I}}^\dagger 
\Delta^\mu(1920) \cdot \partial_\mu \vec\pi \, \right),
\label{pnd}
\end{eqnarray}
\begin{eqnarray}
{\cal L}_{K \Sigma N(1710)} &=&
-ig_{K \Sigma N(1710)}
\left( \bar{N}(1710) \gamma_5 \vec\tau K \cdot \overrightarrow\Sigma 
+ \overrightarrow{\bar \Sigma} \cdot  \bar{K} \vec\tau
\gamma_5 N(1710) \right),
\label{ksb}\\
{\cal L}_{K \Sigma N(1720)} &=&
\frac{g_{K \Sigma N(1720)}}{m_K}
\left( \bar{N}^\mu(1720) \vec\tau (\partial_\mu K) 
\cdot \overrightarrow\Sigma
+ \overrightarrow{\bar \Sigma} \cdot (\partial_\mu \bar{K}) 
\vec\tau N^\mu(1720) \right),
\label{ksc}\\
{\cal L}_{K \Sigma \Delta(1920)} &=&
\frac{g_{K \Sigma \Delta(1920)}}{m_K}
\left( \bar{\Delta}^\mu(1920) \overrightarrow{\cal I}
\cdot \overrightarrow\Sigma \partial_\mu K
+ (\partial_\mu \bar{K}) \overrightarrow{\bar \Sigma} \cdot
{\overrightarrow{\cal I}}^\dagger \Delta^\mu(1920) \right), 
\label{ksd}
\end{eqnarray}

where the ratio of the tensor to the vector coupling constant 
at the ${\rho}NN$ vertex in Eq.~(\ref{rnn}) is used as 
$\kappa{=}f_{{\rho}NN}/g_{{\rho}NN}{=}6.1$~\cite{Bonn}.
The operators $\overrightarrow{\cal I}$ 
and $\overrightarrow{\cal K}$ are defined by 
\begin{eqnarray}
\overrightarrow{\cal I}_{M\mu} &\equiv&  
\displaystyle{\sum_{\ell=\pm1,0}}
(1 \ell \frac{1}{2} \mu | \frac{3}{2} M)
\hat{e}^*_{\ell},\\ 
\overrightarrow{\cal K}_{M M'} &\equiv&  \displaystyle{\sum_{\ell=\pm1,0}}
(1 \ell \frac{3}{2} M' | \frac{3}{2} M)
\hat{e}^*_{\ell},\label{operator}
\end{eqnarray}
where $M$, $\mu$ and $M'$ denote the third components of the 
isospin projections, respectively, while $\vec \tau$ are the 
Pauli matrices.
$N, N(1710), N(1720)$ and $\Delta(1920)$
stand for the nucleon fields 
and the baryon resonances, respectively;
they  are expressed by
$\bar{N}{=}\left( \bar{p}, \bar{n} \right)$
and
$\bar{\Delta}(1920){=}(\bar{\Delta}(1920)^{++},
\bar{\Delta}(1920)^+,\bar{\Delta}(1920)^0,
\bar{\Delta}(1920)^-)$ in isospin space.
The physical representations of the fields are given by:
$K^T{=}\left( K^+, K^0 \right)$,
$\bar{K}{=}\left( K^-, \bar{K^0} \right)$,
$\pi^{\pm}{=}(\pi_1 \mp i \pi_2)/\sqrt{2}$,
$\pi^0{=}\pi_3$ and similarly for the $\rho$-meson fields, and 
$\Sigma^{\pm}{=}(\Sigma_1{\mp}i\Sigma_2)/{\sqrt{2}}$,
$\Sigma^0{=}\Sigma_3$, respectively, 
where the superscript $T$ means the transposition operation.
(Notice the similarity of 
$\vec{\tau} \cdot \vec{\pi}$ and 
$\vec{\tau} \cdot \vec{\Sigma}$.)
The meson fields are defined as either annihilating or creating 
the physical particle or anti-particle states, respectively.

For the propagators, $iS_F(p)$ for spin 1/2 and
$iG^{\mu \nu}(p)$ for spin 3/2 resonances, we use:
\begin{equation}
iS_F(p) = i \frac{\gamma \cdot p + m}{p^2 - m^2 + im\Gamma^{full}}\,,
\label{spin1/2}
\end{equation}
\begin{equation}
iG^{\mu \nu}(p) = i \frac{-P^{\mu \nu}(p)}{p^2 - m^2 +
im\Gamma^{full}}\,,  \label{spin3/2}
\end{equation}
with
\begin{equation}
P^{\mu \nu}(p) = - (\gamma \cdot p + m)
\left[ g^{\mu \nu} - \frac{1}{3} \gamma^\mu \gamma^\nu
- \frac{1}{3 m}( \gamma^\mu p^\nu - \gamma^\nu p^\mu)
- \frac{2}{3 m^2} p^\mu p^\nu \right], \label{pmunu}
\end{equation}
where $m$ and $\Gamma^{full}$ stand for the mass and full decay
width of the corresponding resonances.

To account for the nonlocality or finite size of hadrons 
we introduce form factors, $F_M (\vec{q})$, at the relevant interaction 
vertices,  
\begin{equation}
F_M(\vec{q}) = 
\displaystyle{\left( \frac{\Lambda_M^2}{\Lambda_M^2 
+ \vec{q}\,^2} \right)^n}, 
\label{formfactor}
\end{equation}
with $\vec{q}$ denoting the momentum of the exchanged meson
and $n{=}1$ for the $\pi$ and $\eta$-mesons and $n{=}2$ for  
the $\rho$-meson, respectively, while $\Lambda_M$ denotes 
the cut-off parameter. For the calculation of the $NN{\to}N{\Sigma}K$ 
reactions we adopt the same form factors, coupling constants 
and cut-off parameters which were used in 
Refs.~\cite{Tsushima1,Tsushima2,Sibirtsev3}. 
It should be emphasized that the coupling constants and form factors 
necessary for the calculation of the $NN{\to}N{\Sigma}K$ reactions 
have been already fixed to reproduce the 
${\pi}N{\to}YN$ and $pp{\to}N{\Lambda}K$ total cross section data.
The  coupling constants and cut-off parameters used in the 
calculation are summarized in Table~\ref{tab3}. 

\begin{table}
\caption{\label{tab3}
Coupling constants and cut-off parameters used in the present study.}
\begin{center}
\begin{tabular}{|l|l|c|}
\hline
vertex & $g^2/4\pi$ & cut-off (MeV)\\ 
\hline 
$\pi N N$  & $14.4$ & $1050$ \\ 
$\pi N N(1710)$ &$2.05 \times 10^{-1}$ &$800$ \\
$\pi N N(1720)$ &$4.13 \times 10^{-3}$ &$800$ \\
$\pi N \Delta(1920)$ &$1.13 \times 10^{-1}$ &$500$ \\
$\eta N N$ &$5.00$ &$2000$\\
$\eta N N(1710)$ &$2.31$ &$800$ \\ 
$\eta N N(1720)$ &$1.03 \times 10^{-1}$ &$800$\\
$\rho N N$  &$0.74$ &$920$ \\ 
$\rho N N(1710)$ &$3.61 \times 10^{+1}$ &$800$\\
$\rho N N(1720)$  &$1.43 \times 10^{+2}$ &$800$ \\ 
$K \Sigma N(1710)$ &$4.66$ &$800$ \\
$K \Sigma N(1720)$ &$2.99 \times 10^{-1}$ &$800$\\
$K \Sigma \Delta(1920)$ & $3.08 \times 10^{-1}$ & $500$ \\
\hline
\end{tabular}
\end{center}
\end{table}

\section{Analysis of model uncertainties}
In this section we analyze the contributions from different 
meson exchanges  and the sensitivity to the model parameters.

     From a first look at the coupling constants listed in 
Table~\ref{tab3} one expects that the dominant contribution for 
the $NN{\to}N{\Sigma}K$ reaction 
should come from the $N(1710)$  or $P_{11}$ resonance. 
The $g_{K{\Sigma}N(1710)}$ coupling constant is one order 
of magnitude larger than  
$g_{K{\Sigma}N(1720)}$ and $g_{K{\Sigma}\Delta(1920)}$.  
Although the $g_{{\rho}NN(1720)}$ coupling constant is larger than 
$g_{{\rho}NN(1710)}$, it is compensated by the small value for 
$g_{K{\Sigma}N(1720)}$ coupling constant. 
The actual calculated results support  
this simple estimate. 

\begin{figure}[ht]
\psfig{file=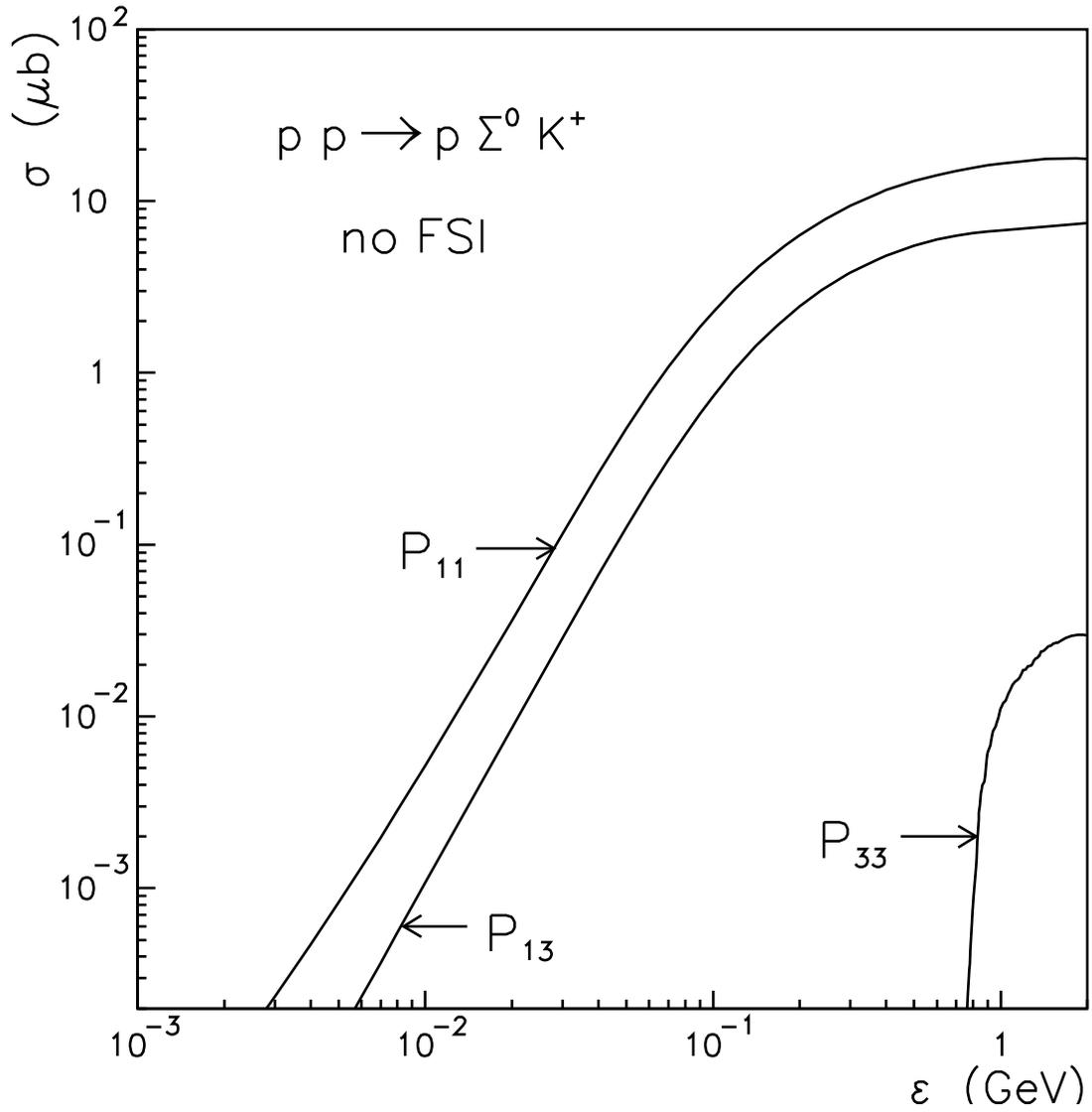,width=16cm}
\caption[]{\label{kn4} Decomposition of the total cross section 
in terms of individual resonance contributions, $N(1710)$ ($P_{11}$),
$N(1720)$ ($P_{13}$) and $\Delta(1920)$ ($P_{33}$) resonances for the  
$pp{\to}p\Sigma^0K^+$ reaction. The calculation was performed 
without the inclusion of final state interactions.}
\end{figure}

In Fig.~\ref{kn4} we show the decomposition of the contributions 
from each resonance to the energy dependence of the 
$pp{\to}p\Sigma^0K^+$ total cross section, plotted as a function of
the excess energy $\epsilon{=}\sqrt{s}{-}m_N{-}m_\Sigma{-}m_K$.
The calculations were performed without including the 
${\Sigma}N$ final state interaction (FSI), which is expected to 
be important at energies very close to 
threshold~\cite{Balewski3,Sibirtsev4}. 
Indeed, the results shown in Fig.~\ref{kn4} confirm the 
expectation that the dominant contribution comes from the 
$P_{11}$ resonance. In addition, the contribution from the 
$P_{33}$ baryonic resonance is almost negligible. 

\begin{figure}[ht]
\psfig{file=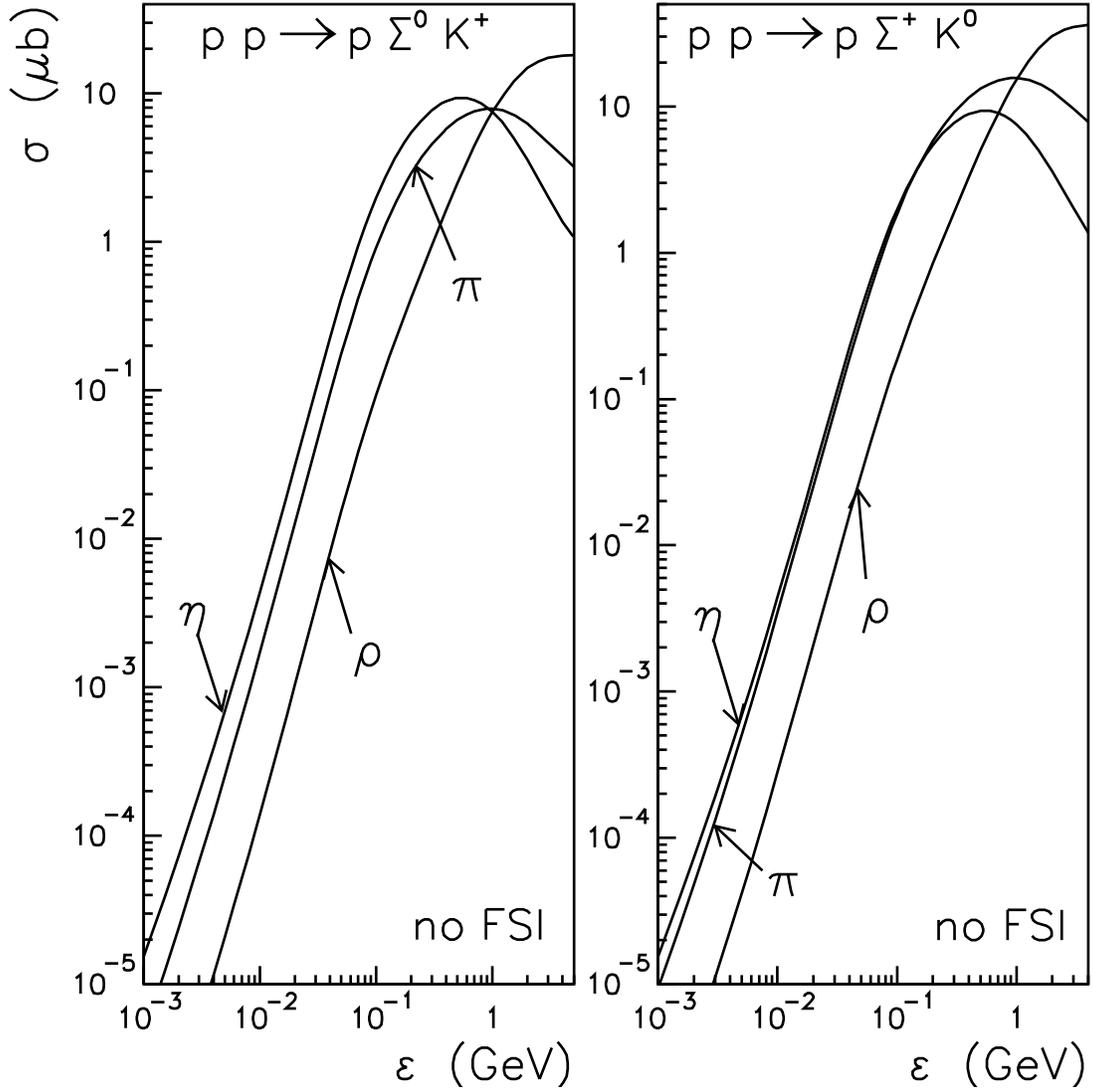,width=16cm}
\caption[]{\label{kn1} Decomposition of the total cross section 
in terms of individual  meson exchange contributions, $\pi$, $\eta$ and 
$\rho$-meson exchanges for the $pp{\to}p\Sigma^0K^+$ and 
$pp{\to}p\Sigma^+K^0$ reactions using $g_{{\eta}NN}{=}7.9$.}
\end{figure}

Next, we analyze the contributions from the different  
meson exchanges. It is not straightforward to estimate, which 
meson exchange gives a dominant contribution to the ${\Sigma}K$ 
production in nucleon-nucleon collisions by just looking at 
the data in Table~\ref{tab3}.   
We, therefore, show in Fig.~\ref{kn1} the separate contributions 
from $\pi$, $\eta$ and $\rho$-meson exchanges to the 
energy dependence of the $pp{\to}p\Sigma^0K^+$ and 
$pp{\to}p\Sigma^+K^0$ total cross sections.  
We find that the contributions from $\pi$ and $\eta$ exchange 
are important
at low energies, while the $\rho$-meson exchange contribution 
dominates at excess energies $\epsilon$ above 1~GeV.

\begin{figure}[ht]
\psfig{file=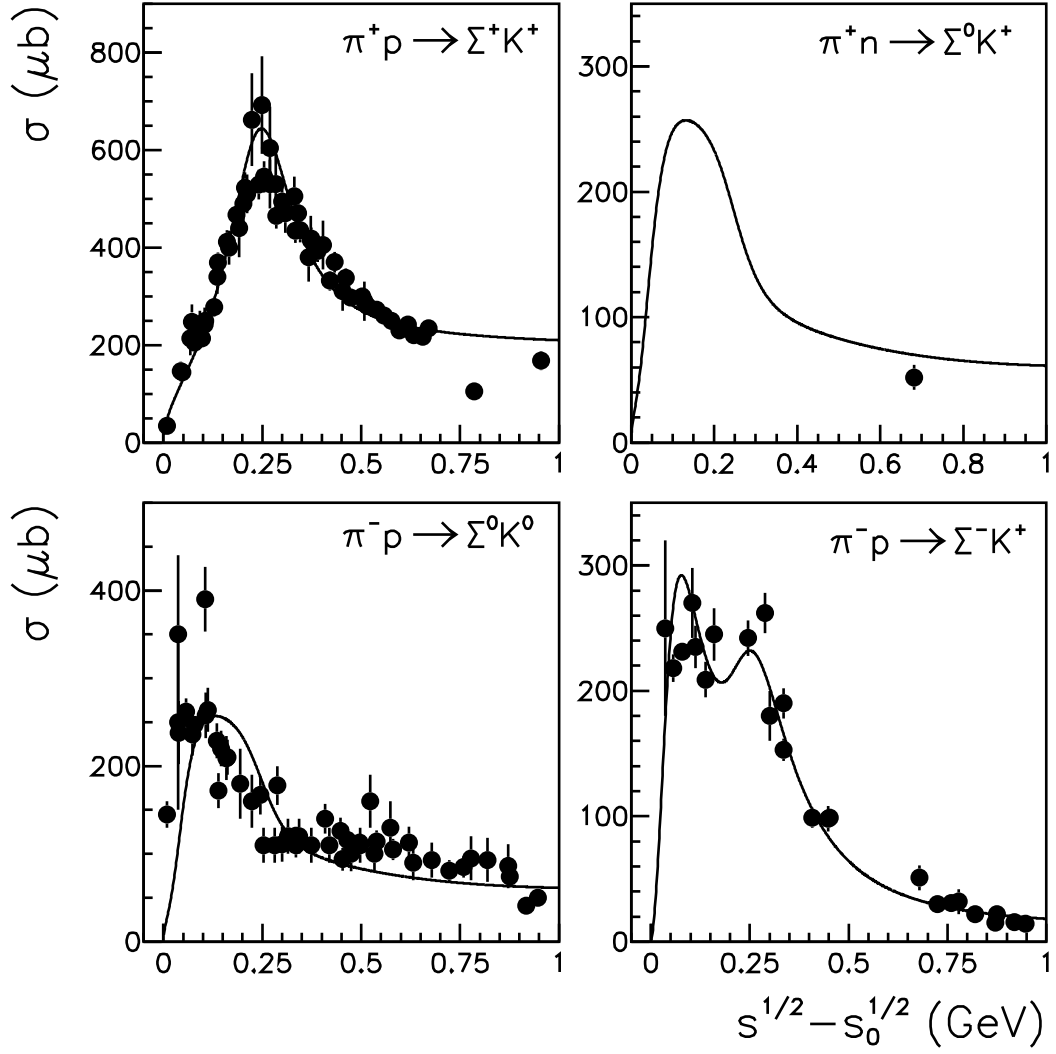,width=16cm}
\caption[]{\label{kn14} Energy dependence of the total cross sections
for the $\pi^+p{\to}K^+\Sigma^+$,  $\pi^-p{\to}K^0\Sigma^0$,
$\pi^+n{\to}K^+\Sigma^0$ and  $\pi^-p{\to}K^+\Sigma^-$
reactions as a function of the excess energy above the reaction 
threshold. The experimental data (full dots with error bars) are 
taken from Ref.~\protect\cite{LB} while the solid lines show 
our calculations.}
\end{figure}

A crucial question is the sensitivity of the results to the 
coupling constants and cut-off parameters which cannot be 
fixed solely from the experimental data within the model. 
The situation for the $\pi$-meson exchange is rather clear;
the coupling constant and the form factor at the ${\pi}NN$ 
vertex are well under control from the study of $NN$ interactions. 
In addition, according to the investigations of the ${\pi}N{\to}{\Sigma}K$
reactions~\cite{Tsushima1}, the ${\pi}NR$ and $K{\Sigma}R$ vertices
are fixed consistently by the existing experimental data. 
Fig.~\ref{kn14} shows a comparison between the model calculations
and the data on ${\pi}N{\to}{\Sigma}K$ cross sections. With the
set of the parameters listed in Tab.~\ref{tab2} the model
reasonably reproduces the different reaction channels:
$\pi^+p{\to}K^+\Sigma^+$,  $\pi^-p{\to}K^0\Sigma^0$,
$\pi^+n{\to}K^+\Sigma^0$ and  $\pi^-p{\to}K^+\Sigma^-$.

On the other hand, the situation for the $\eta$-meson exchange 
is less obvious.  Since the contribution from $\eta$-exchange  
is important at low energies (cf. Fig.~\ref{kn1}),  
we have to discuss the status of the ${\eta}NN$ coupling constant 
in more detail.

As reviewed by Tiator, Benhold and Kamalov~\cite{Tiator},  
the ${\eta}NN$ coupling constant 
has still large uncertainties. The values for $g_{{\eta}NN}$ vary 
between 1 and 9 and depend substantially on the model 
applied to describe the data. Furthermore, the experimental data 
themselves have large uncertainties and do not allow 
to extract a precise 
value for $g_{{\eta}NN}$. 
When using the one-boson exchange potential to fit the nucleon-nucleon
phase shifts~\cite{Brockmann} one requires   
$6{\le}g_{{\eta}NN}{\le}9$. However, the analysis itself is not 
sensitive enough to the $\eta$-meson exchange. 
On the other hand, the $\pi^-p{\to}{\eta}n$
data restrict the value to be small, 
$g_{{\eta}NN}{=}2.7{\div}4.6$~\cite{Tiator}.
 
Furthermore, a fit to the total cross section for 
the ${\gamma}p{\to}{\eta}p$ reaction requires 
$g_{\eta NN}{=}1.1$ in the case of pseudoscalar coupling and 
$f_{\eta NN}{=}8.7$ for pseudovector coupling~\cite{Tiator}.  
Thus the total cross section is not sensitive enough to 
evaluate the coupling constant. However, the authors of 
Ref.~\cite{Tiator} suggest  a value, $g_{{\eta}NN}{=}2.24$, 
which is extracted from the description of the angular distribution. 
A global analysis of both meson scattering and photon induced 
reactions on the nucleon, performed recently by Feuster and 
Mosel~\cite{Feuster}, yields a value  $g_{{\eta}NN}{=}1.0$. 

Based on SU(3) symmetry the $g_{{\eta}NN}$ coupling constant is related 
to the $g_{{\pi}NN}$ as~\cite{Carruthers} 
\begin{equation}
g_{{\eta}NN}=\frac{-\alpha + 3(1- \alpha)}{\sqrt{3}} \ g_{{\pi}NN}
\end{equation}
with $\alpha{=}0.6$ when combined with the SU(6) quark model; 
this value for $\alpha$ gives $g_{{\eta}NN}{\simeq}4.65$. 

In our calculation we use the value $g_{{\eta}NN}{=}7.9$ 
taken from the Bonn $NN$ potential model~\cite{Bonn}.
Considering the uncertainty discussed above for the ${\eta}NN$ 
coupling constant,  
we have investigated how the calculated results are affected 
by varying the value for $g_{{\eta}NN}$. 
The deviations due to the different values for $g_{{\eta}NN}$ for the 
contribution from the $\eta$-exchange ( shown in Fig.~\ref{kn1}
for $g_{{\eta}NN}{=}7.9$ ) 
then vary by factors of 1/60 to 1.3, respectively.

In addition, there are also uncertainties in the resonance properties.
The latest issue of the Review of Particle Data~\cite{PDG1}
no longer lists the branching ratios for the 
$N(1710){\to}{\eta}N$ and  $N(1710){\to}{\rho}N$ channels. 
Thus, the coupling constants $g_{{\eta}NN(1710)}$ and 
$g_{{\rho}NN(1710)}$ also involve uncertainties on the present 
experimental confidence level. A similar situation also holds for the 
$N(1720)$ resonance. Note, however, that the $\rho$ couplings 
have been already fixed
by our previous study on $\Lambda$-hyperon production in $NN$
collisions~\cite{Tsushima2}.

\section{Comparison with data}

\begin{figure}[ht]
\psfig{file=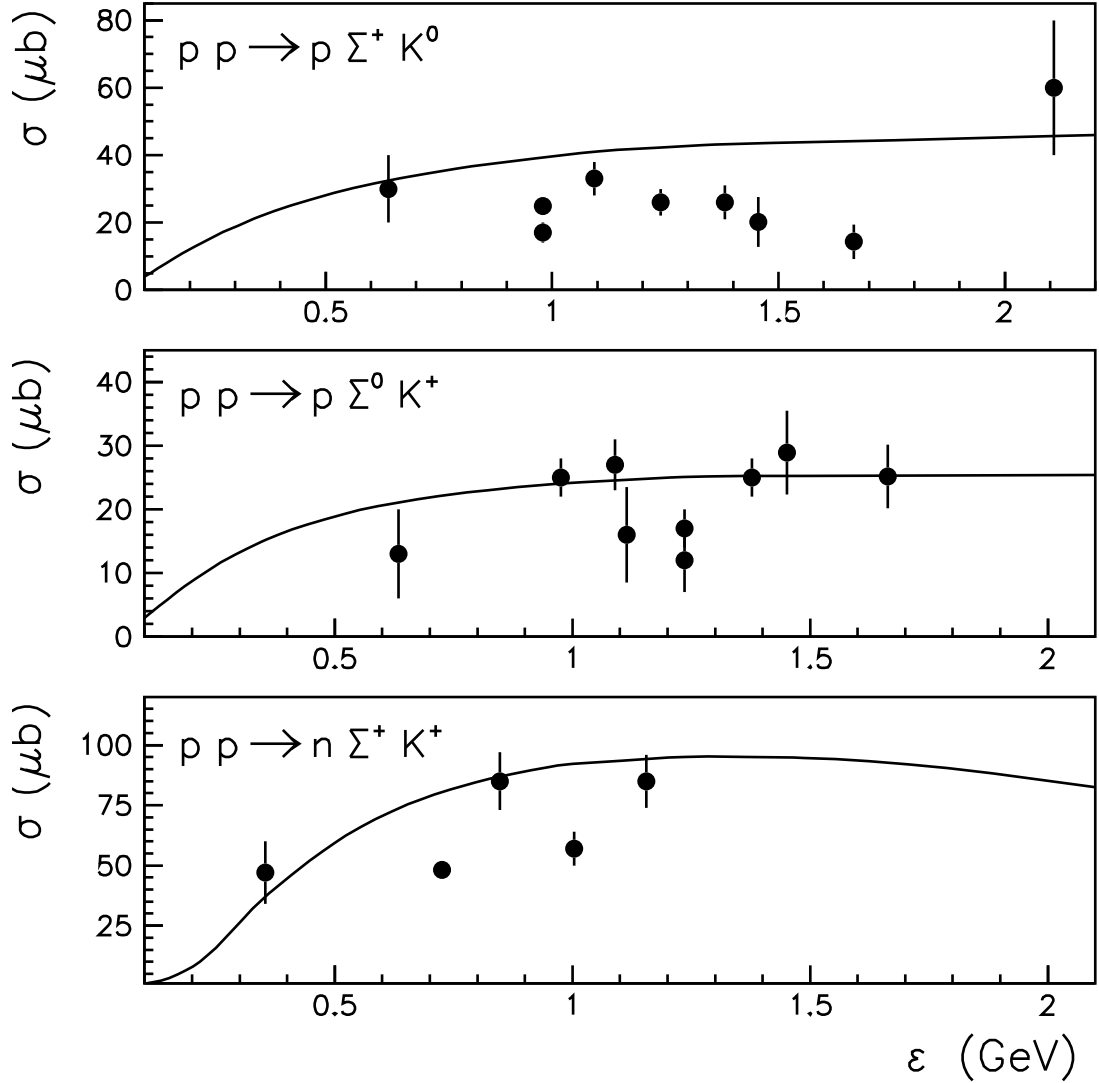,width=16cm}
\caption[]{\label{kn9} Energy dependence of the total cross sections 
for the $pp{\to}p\Sigma^+K^0$, $pp{\to}p\Sigma^0K^+$ and
$pp{\to}n\Sigma^+K^+$ reactions as a function of the
excess energy $\epsilon$ above the reaction threshold. 
The experimental data (full dots with error bars) are taken from 
Ref.~\protect\cite{LB} while the solid lines show the calculations 
performed without the inclusion of final state interactions.} 
\end{figure}

The data for $NN{\to}N{\Sigma}K$ reactions are available
at excess energies larger than 300~MeV. In Fig.~\ref{kn9}
we show the calculated energy dependence of the total cross sections for 
the $pp{\to}p\Sigma^+K^0$, $pp{\to}p\Sigma^0K^+$ 
and $pp{\to}n\Sigma^+K^+$ reactions (solid lines) in comparison 
with the experimental data (dots) from Ref.~\cite{LB}.
Our model reproduces all these reaction channels reasonably 
well with the same set of parameters.

\begin{figure}[ht]
\psfig{file=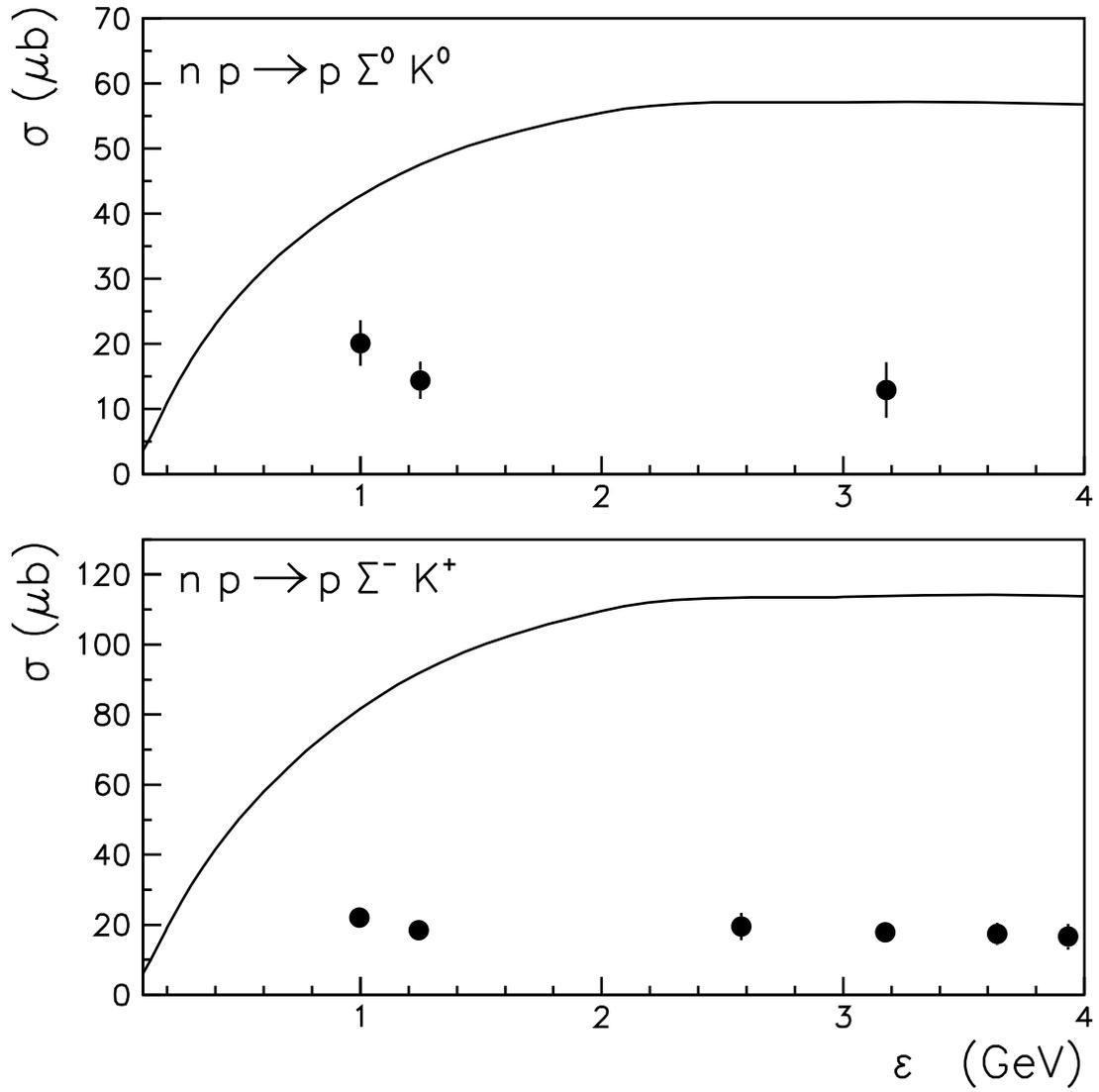,width=16cm}
\caption[]{\label{kn10} Energy dependence of the total cross sections 
for the $np{\to}p\Sigma^0K^0$ and
$np{\to}p\Sigma^-K^+$ reactions; same notation as in 
Fig.~\protect\ref{kn9}.} 
\end{figure}

\begin{figure}[ht]
\psfig{file=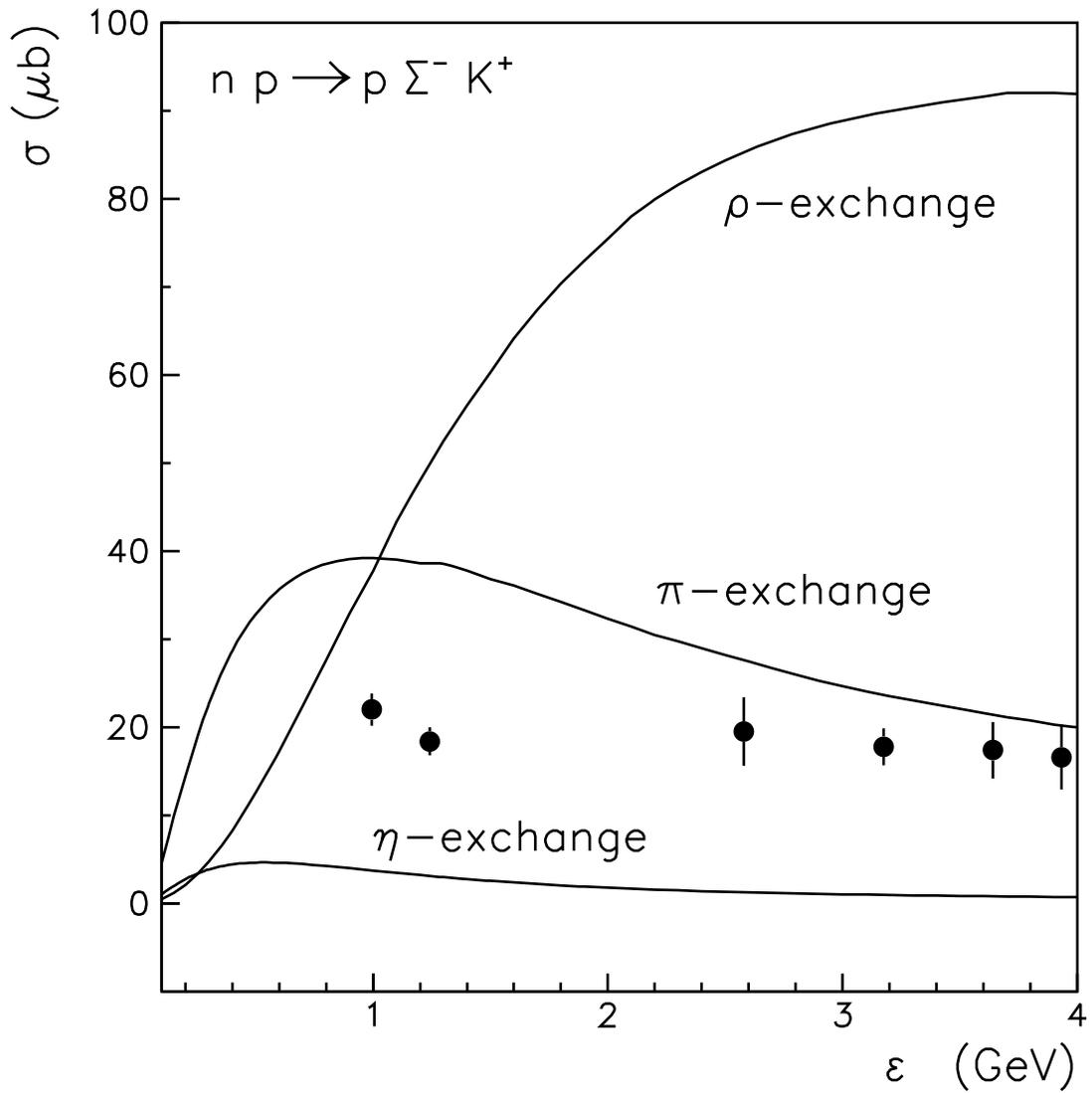,width=16cm}
\caption[]{\label{kn11} Decomposition of the total cross section
in terms of meson exchange contributions, $\pi$, $\eta$ and
$\rho$-meson exchanges, for the $np{\to}p\Sigma^-K^+$ reaction.}
\end{figure}

However, the situation is different for the
$np{\to}p\Sigma^0K^0$ and $np{\to}p\Sigma^-K^+$ reactions. 
The energy dependence of the total cross sections for these 
reactions is shown in Fig.~\ref{kn10}; here
our calculations (solid lines) substantially overestimate
the experimental data (full dots) from Ref. \cite{LB}. 
As illustrated in Fig.~\ref{kn1},
at excess energies above 1~GeV the dominant contribution
stems from $\rho$-meson exchange according to our model. 
This phenomenon is shown in Fig.~\ref{kn11}, where the meson exchange
contributions are displayed separately. 
This implies that almost all available data for the 
$NN{\to}N{\Sigma}K$ reaction are dominantly governed by the 
parameters for $\rho$-exchange, while  
the model parameters, together with those for $\rho$-exchange, 
have been fixed to reproduce optimally 
the existing total cross section data for the  
$pp{\to}p{\Lambda}K^+$ and $pp{\to}N{\Sigma}K$ reactions 
at excess energies around 1 GeV.  Thus the data 
for the $np{\to}p{\Sigma}K$ reaction at excess energies larger 
than 1 GeV clearly indicate the limitations of our model. 

\begin{figure}[ht]
\psfig{file=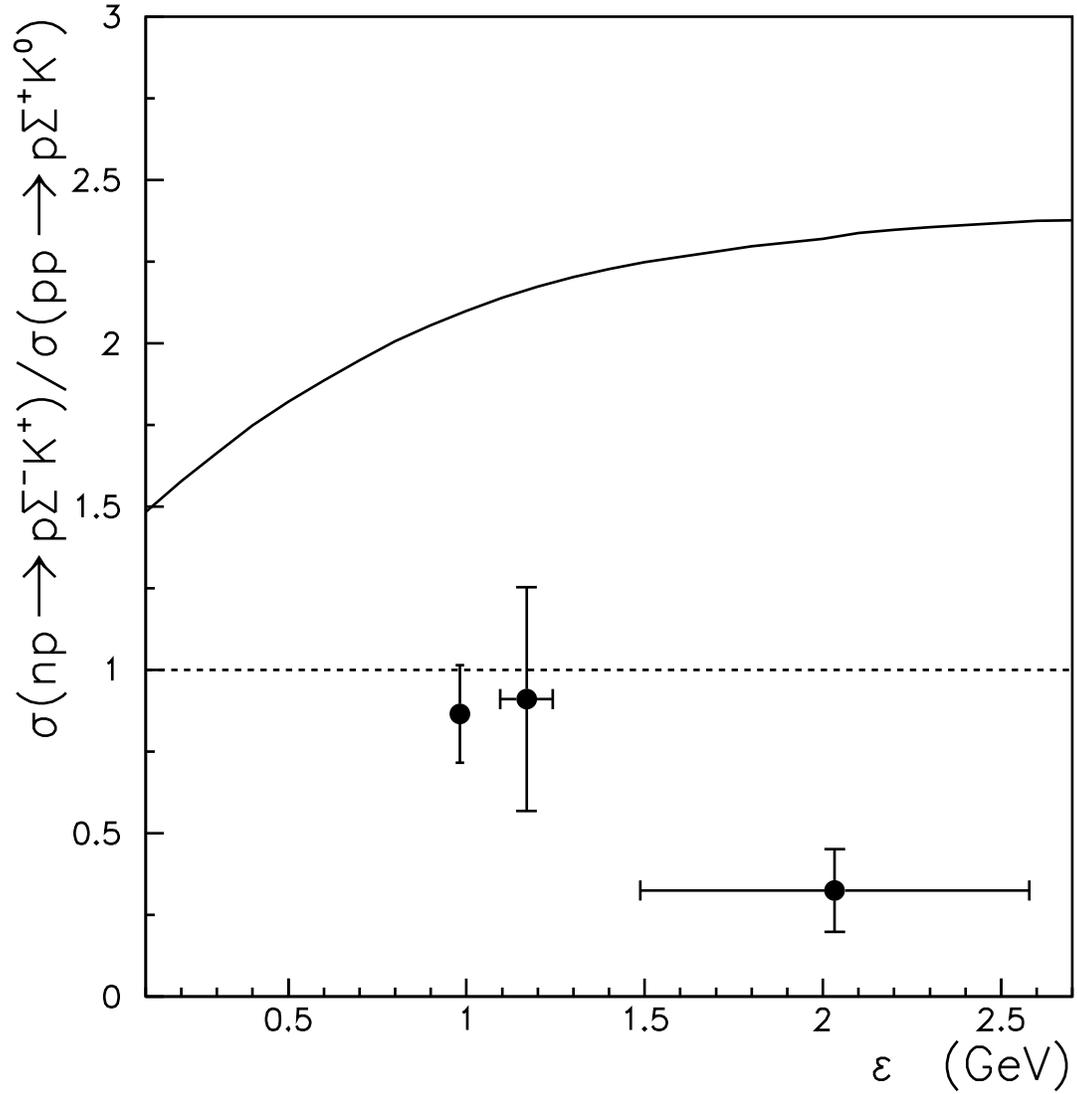,width=16cm}
\caption[]{\label{kn12} Ratios of the $np{\to}p\Sigma^-K^+$ total 
cross section to that for $pp{\to}p\Sigma^+K^0$. 
The ratios for the experimental data (full dots with error bars) 
are taken from Ref.~\protect\cite{LB}, while the solid line shows
the ratio for the calculated results.}
\end{figure}

Nevertheless, we may check the consistency of the data.
For this purpose we show the ratios of the total cross sections 
for the $np{\to}p\Sigma^-K^+$ to $pp{\to}p\Sigma^+K^0$ both 
for the experimental data (full dots with error bars) and 
the calculated results (solid line) in Fig.~\ref{kn12}. 
In a simple one-boson exchange picture (tree-level),   
only neutral mesons can be exchanged in the 
$pp{\to}p\Sigma^+K^0$ reaction, while both neutral and charged 
mesons can be exchanged in the $np{\to}p\Sigma^-K^+$
reaction. Furthermore, the isospin factor for charged meson exchange 
at the meson-nucleon-nucleon vertex is a factor of $\sqrt{2}$ 
larger than that for the neutral meson.  
If we neglect the small contribution from $\eta$ exchange 
as in Fig.~\ref{kn11}, our explicit calculation shows that the direct 
amplitudes for $\pi$ and $\rho$ exchange followed by  
$N(1710)$ and $N(1720)$ resonance excitation, give a factor of 
two larger contributions to the $np{\to}p\Sigma^-K^+$ reaction 
relative to those for the $pp{\to}p\Sigma^+K^0$ reaction.   
Thus, we may naively expect that the total cross section for the 
$np{\to}p\Sigma^-K^+$ reaction should be larger than that for 
$pp{\to}p\Sigma^+K^0$.  

However, the ratios for the experimental data in 
Fig.~\ref{kn12} (the dots) clearly contradict this expectation; 
the ratios become even smaller than one. The 
calculated results (solid line) reasonably 
agree with the simple estimate made above.  
We note that the available data for the $\Sigma K$ production in $np$
collisions collected in Ref.~\cite{LB} were obtained by 
the group of Ansorge et al.~\cite{Ansorge} alone. The data were taken 
with a neutron beam and the CERN 2~m hydrogen bubble chamber;  
the systematic uncertainty due to the beam normalization  
$\simeq$13\% is not shown in our figures. 
Thus, we would like to urge experimental collegues to collect more 
data for the $np{\to}N{\Sigma}K$ reaction in order to resolve
this systematic discrepancy. 

\begin{figure}[ht]
\psfig{file=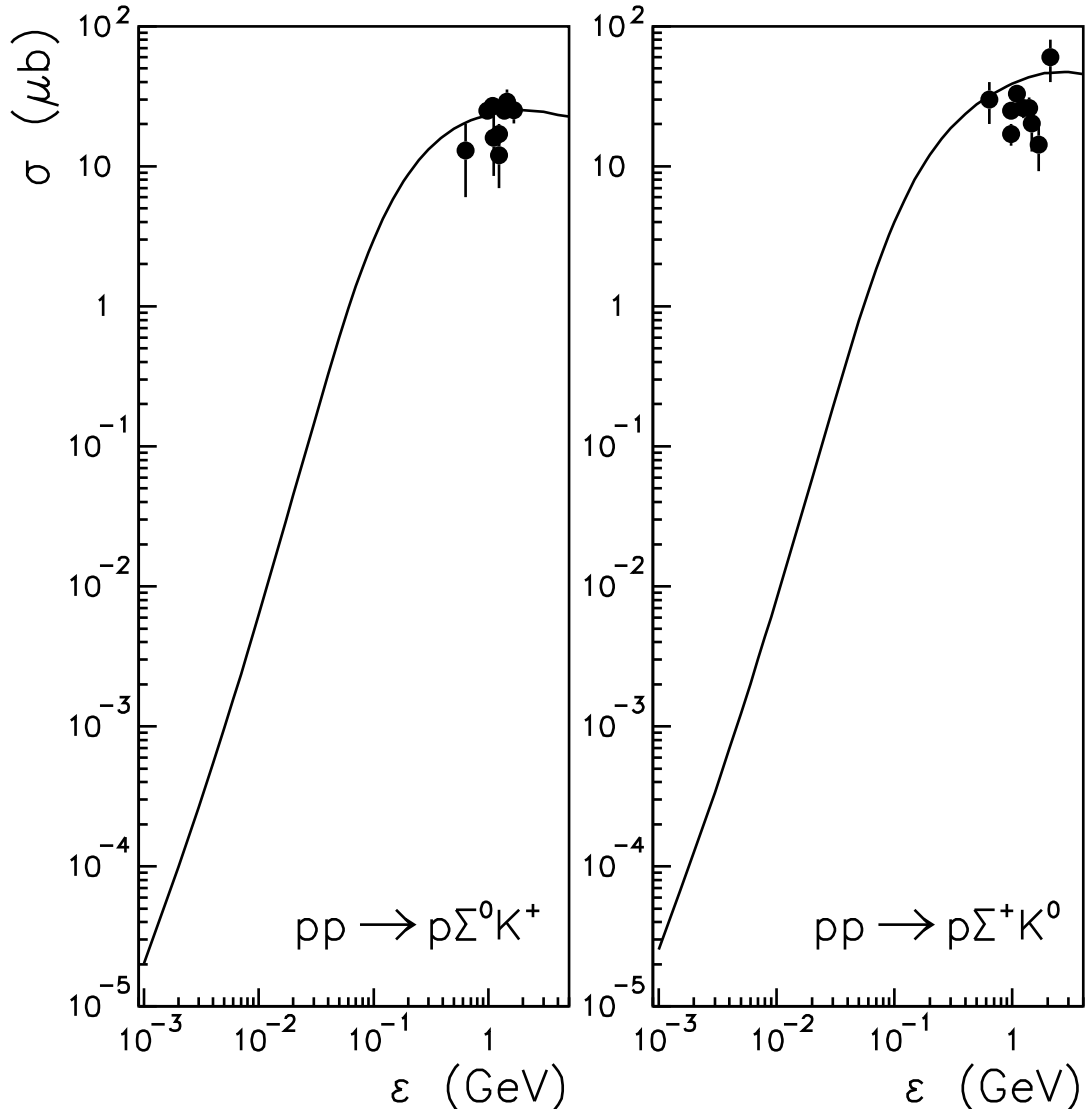,width=16cm}
\caption[]{\label{kn2} Energy dependence of the total cross sections 
for the $pp{\to}p{\Sigma^0}K^+$ and $pp{\to}p{\Sigma^+}K^0$ reactions  
as a function of the excess energy 
$\epsilon$  above the reaction threshold. 
Note that the final state interaction is not included; it is expected 
to be important at energies very close to the threshold.}
\end{figure}

Finally, in Fig.~\ref{kn2} we show the $pp{\to}p\Sigma^0K^+$ 
total cross section for a larger scale of 
excess energies, as relevant for COSY, such that one can 
examine the effect of final state interactions in more detail. Note, that 
our calculations were performed without the inclusion of 
FSI. Thus we expect that the present results 
should deviate from the experimental data at small excess energies. 
Deviations of our calculation~\cite{Tsushima2} 
from the data~\cite{Balewski1,Balewski2} at excess 
energies $\epsilon{\le}10$~MeV have been also noticed for 
the $pp{\to}p{\Lambda}K^+$ reaction. It was argued both
experimentally~\cite{Balewski3} and theoretically~\cite{Sibirtsev4}, 
that at energies very close to the reaction threshold the production 
mechanism is strongly distorted by the final state interaction 
between the hyperon and the nucleon. 

One of the common ways~\cite{Sibirtsev5,Shyam,Moalem,Bernard} 
to account for FSI is using the Watson-Migdal 
theorem~\cite{Watson,Migdal}, which expresses the total reaction 
amplitude as a product of the production 
amplitude and the on mass-shell $YN{\to}YN$ scattering amplitude.
However, this approach was recently criticized 
by Hanhart and Nakayama~\cite{Hanhart} focussing on the on-shell 
treatment and factorization of the reaction amplitude.
It was also suggested by Mei{\ss}ner~\cite{Meissner} that the
treatment of FSI might be too simplistic in the Watson-Migdal scheme. 

In addition, we note that the knowledge of the $YN$ scattering
amplitudes is still unsatisfactory. Although there have been extensive 
theoretical studies of the hyperon-nucleon 
interaction~\cite{Nagels,Takahashi,Maessen,Holzenkamp,Reuber,Rijken},
the $YN$ interaction models from Nijmegen~\cite{Maessen}
and J\"ulich~\cite{Holzenkamp} predict completely different
results  for the $pn{\to}{\Lambda}p$ reaction  
as recently shown by Parreno et al.~\cite{Parreno}. 
This implies that the introduction of FSI corrections  in 
the calculation is 
strongly model dependent even within the on mass-shell
Watson-Migdal approximation. The situation is even more 
complicated~\cite{Laget} because the 
${\Sigma}N{\leftrightarrow}{\Lambda}N$ channel coupling 
becomes stronger at energies close to the threshold. 
It is therefore not so easy to improve the 
$NN{\to}NYK$ calculations near threshold by taking into account FSI
corrections in an unambiguous manner, as compared to the treatment 
proposed in Ref.~\cite{Sibirtsev4}. As a consequence, we discard the 
explicit calculation
of FSI corrections and rather concentrate  
our study on excess energies larger than 40~MeV, 
where FSI and the ${\Sigma}N{\leftrightarrow}{\Lambda}N$ channel coupling 
are expected to be small.  

\section{Analysis of the reaction dynamics within the model}

To obtain insight into the dynamics of hyperon production in
$NN$ collisions we demonstrated in Ref.~\cite{Sibirtsev3} that 
the invariant mass spectrum for the ${\Lambda}K$ system
indeed provides a clue to the $\Lambda$ production mechanism.
The situation also holds true for the $NN{\to}N{\Sigma}K$ reaction.
If the $NN{\to}N{\Sigma}K$ reaction proceeds through the
excitation of baryonic resonances one might observe 
a characteristic distribution for the ${\Sigma}K$
invariant mass, which reflects the excitation of the
intermediate baryonic resonance. As shown in Ref.~\cite{Sibirtsev2} the
invariant mass distribution for the hyperon-nucleon system
extends to the upper limit of $m_Y{+}m_K{+}\epsilon$ and, 
thus, depends on the excess energy $\epsilon$. Since the typical 
widths of the baryonic resonances (cf. Table~\ref{tab2})  
range from 100 to 200~MeV, the distribution may 
be affected by intermediate baryonic resonance excitations
for $\epsilon{\ge}100$~MeV (half width of 200~MeV).
This also implies that the invariant mass spectrum at energies 
near the reaction threshold, i.e., small excess energies, 
will not show a clear signal 
of the baryonic resonance excitation.  Since the invariant 
mass distribution ranges from the threshold energy to the 
threshold energy plus excess energy, the shape of the invariant mass 
distribution cannot be affected noticeably if the excess energy is 
smaller than the resonance width.

\begin{figure}[ht]
\psfig{file=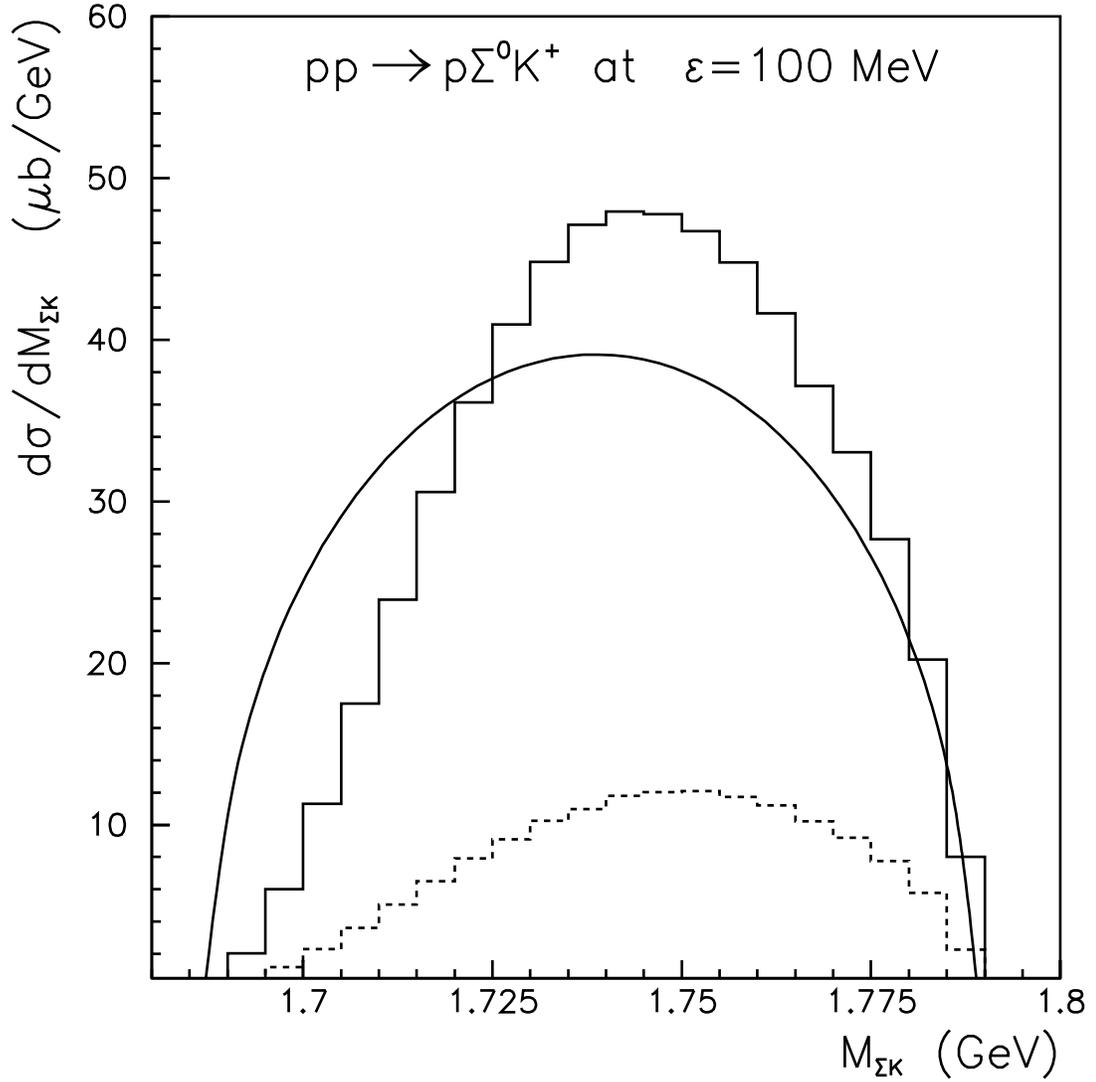,width=16cm}
\caption[]{\label{kn5} Calculated invariant mass distribution
for the ${\Sigma}K$ system produced in the $pp{\to}p\Sigma^0K^+$ 
reaction at an excess energy of 100~MeV. The dashed histogram shows 
the sum of the contribution from the $N(1720)$ and ${\Delta}(1920)$ 
resonances, while the solid histogram shows the total contribution 
from the $N(1710)$, $N(1720)$ and ${\Delta}(1920)$ resonances. 
The solid line is a phase-space distribution
normalized to the same total cross section.}
\end{figure}

In Fig.~\ref{kn5} we show the ${\Sigma}K$ invariant mass 
distribution for the $pp{\to}p\Sigma^0K^+$ reaction calculated 
at an excess energy of 100~MeV.
The dashed histogram is the sum of the contributions from
the $N(1720)$ and $\Delta{(1920)}$ resonances, while the 
solid histogram corresponds to the total sum of contributions from 
the $N(1710)$, $N(1720)$ and $\Delta{(1920)}$ resonances.
It is clear that the dominant contribution comes from the $N(1710)$. 
For comparison, we also show the phase space distribution 
normalized to the same total cross section 
(solid line). At this excess energy the deviation from 
the phase-space distribution due to resonance excitations 
is already noticeable.

\begin{figure}[ht]
\psfig{file=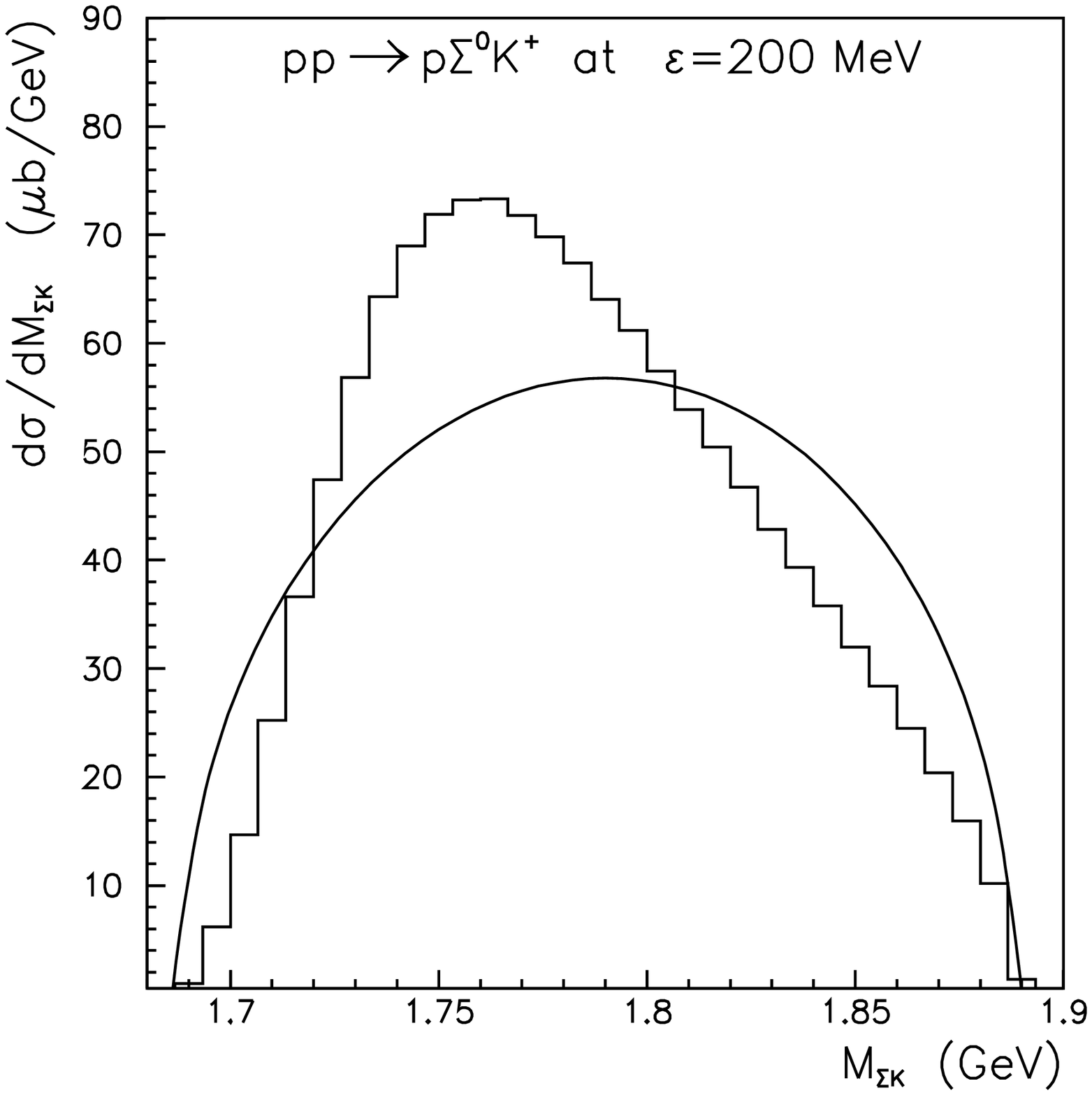,width=16cm}
\caption[]{\label{kn6} Same as  Fig.~\protect\ref{kn5} 
 at an excess energy of 200~MeV.} 
\end{figure}

In Fig.~\ref{kn6} we show again the ${\Sigma}K$ invariant mass 
distribution for the $pp{\to}p\Sigma^0K^+$ reaction, but 
at an excess energy $\epsilon{=}$200~MeV. The solid histogram 
is the sum of the contributions 
from all the resonances, $N(1710)$, $N(1720)$ and $\Delta(1920)$,
while the solid line is the phase-space distribution
normalized to the same total cross section. 
At this excess energy the deviation from the phase-space distribution 
becomes more pronounced and the signal from the $N(1710)$ excitation can
be detected at the peak position of the $\Sigma K$ invariant mass, 
$M_{\Sigma K}{\simeq}1.76$~GeV. Recall, that the contributions from 
the other resonances are small.

\begin{figure}[ht]
\psfig{file=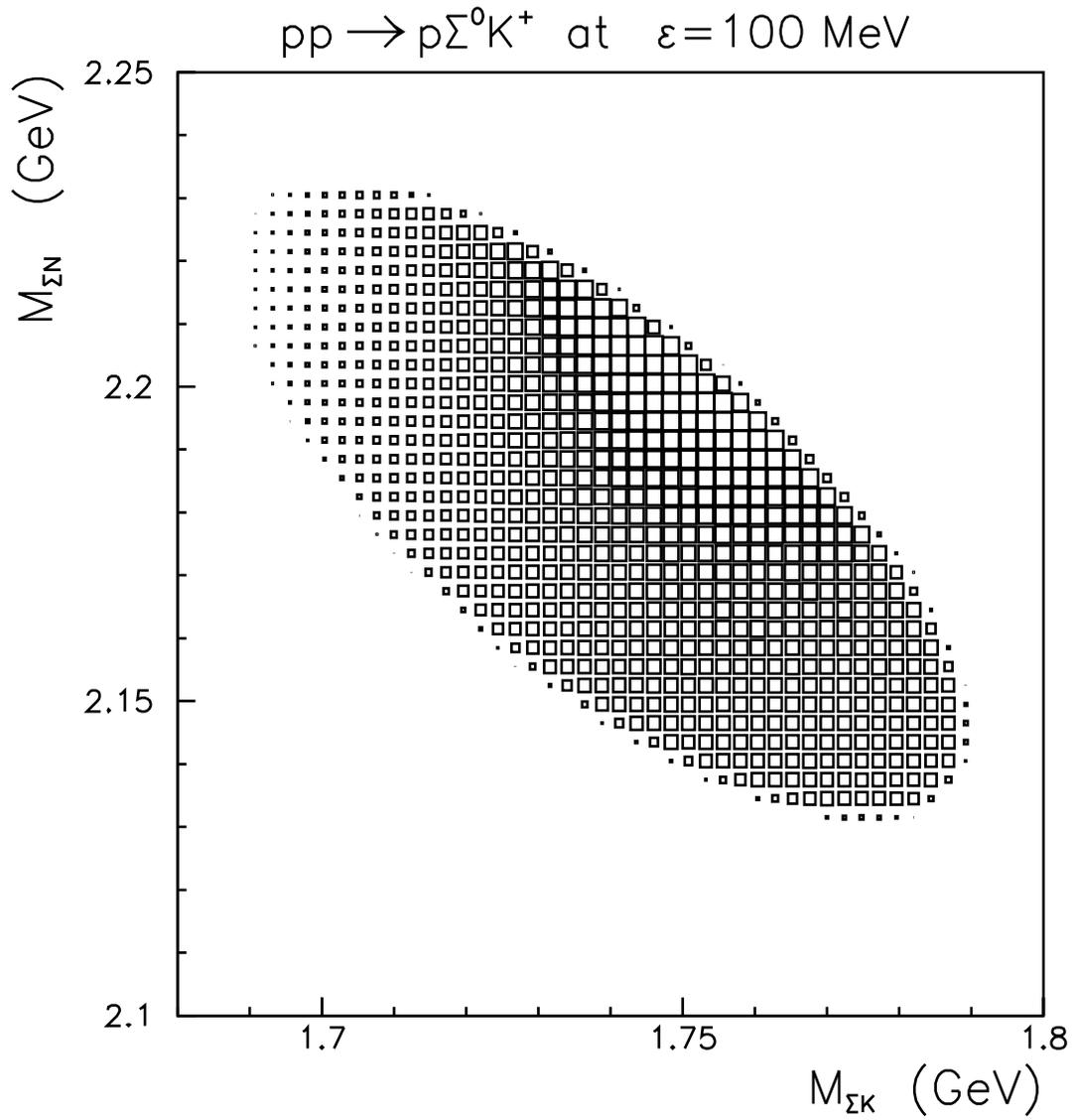,width=16cm}
\caption[]{\label{kn7} Dalitz plot for the $pp{\to}p\Sigma^0K^+$ 
reaction versus $M_{\Sigma N}$ and $M_{\Sigma K}$ 
at an excess energy of 100~MeV. The larger squares stand for a 
higher density.}
\end{figure}

As was proposed by the COSY-11~\cite{Balewski3} and TOF-~\cite{TOF} 
Collaborations a more precise analysis can be performed
in terms of a Dalitz plot to understand the ${\Sigma}K$ 
production mechanism. 
In Fig.~\ref{kn7} we show the Dalitz
plot calculated for the $pp{\to}p\Sigma^0K^+$ reaction 
at the excess energy $\epsilon{=}$100~MeV; i.e. the ${\Sigma}N$ invariant
mass versus the ${\Sigma}K$ invariant mass. In Fig.~\ref{kn7}
the larger squares denote a higher density of the invariant 
mass distribution. In case of strong final state interactions between the
$\Sigma$ and nucleon  one would expect an 
enhancement in the lower region of the ${\Sigma}N$ invariant mass. 
According to the hyperon-nucleon potential models of the
Nijmegen~\cite{Maessen,Rijken} and J\"ulich~\cite{Holzenkamp,Reuber} 
groups
the hyperon-nucleon interaction becomes very strong at the  
${\Sigma}N$ invariant mass below $\simeq$2.17~GeV. 
Thus, FSI can substantially distort the $N(1710)$ signal.
Indeed, Fig.~\ref{kn7} illustrates that there is a structure 
around $M_{{\Sigma}K}{=}1.75$~GeV, but this is due to  
the contribution from the $N(1710)$ resonance which overlaps
with the $M_{{\Sigma}N}$ region, where the influence of
FSI might be substantial (cf. Fig.~\ref{kn5}).

\begin{figure}[ht]
\psfig{file=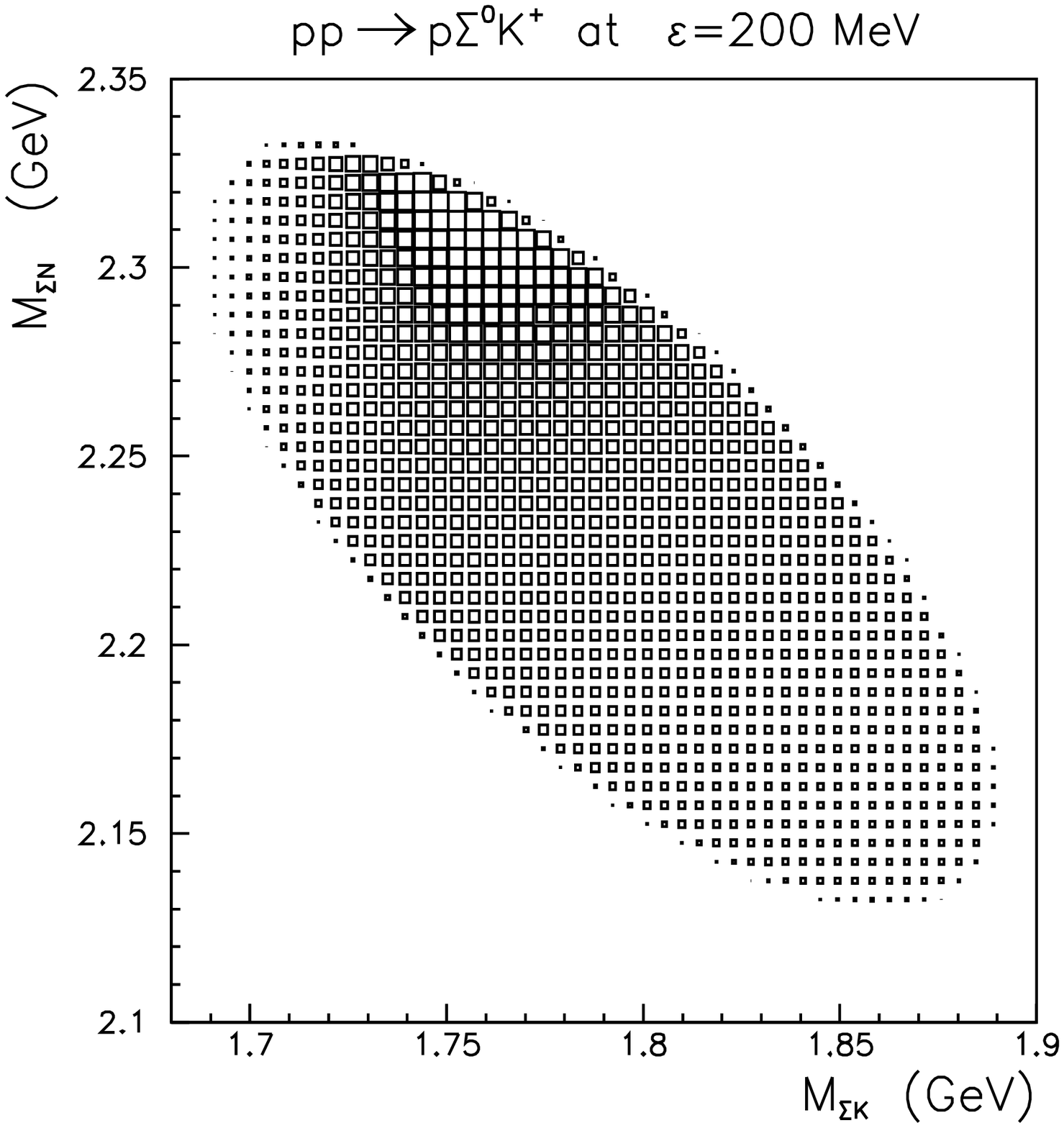,width=16cm}
\caption[]{\label{kn8} Same as  Fig.~\protect\ref{kn7} 
at an excess energy of 200~MeV.}
\end{figure}

In Fig.~\ref{kn8} we show the Dalitz plot for the $pp{\to}p\Sigma^0K^+$ 
reaction at an excess energy $\epsilon{=}200$~MeV.
The situation is much clearer than in the case 
$\epsilon = 100$~MeV, shown in Fig.~\ref{kn7}, since the 
structure due to the $N(1710)$ resonance is well separated 
from the lower range of ${\Sigma}N$  invariant mass. 
Thus, we expect that the Dalitz plots for the 
$M_{{\Sigma}K}$ and $M_{{\Sigma}N}$ invariant masses 
should be a very useful tool for understanding  
the dynamics of the $NN{\to}N{\Sigma}K$ reaction.

\section{Summary}

Within a simple model of one-boson exchange followed by   
baryonic resonance excitation we have investigated the
$NN{\to}N{\Sigma}K$ reactions. 
Our model includes only those mesons which are observed to 
couple to the baryonic resonances which 
themselves are known to decay 
to a $\Sigma$-hyperon and $K$-meson. Thus only the nonstrange meson 
exchanges ($\pi$, $\eta$ and $\rho$) were taken into account.

We demonstrated that our model could reproduce reasonably well 
the available total cross section data for 
the $pp{\to}N{\Sigma}K$ reaction, while it overestimated systematically 
the data for the channel $np{\to}N{\Sigma}K$ at higher energy. 
Concerning the experimental 
data for the  $np{\to}N{\Sigma}K$ reaction, we investigated 
the data consistency in terms of a simple isospin model using the 
one-boson exchange picture. We demonstrated, that it is essential 
to have new experimental data on  neutron induced $\Sigma$ production. 

We also showed that the ${\Sigma}K$ invariant 
mass spectrum provides a powerful method to understand the 
$NN{\to}N{\Sigma}K$ reaction mechanism. 
In the present approach the dominant contribution
to $NN{\to}N{\Sigma}K$ reactions comes from the $N(1710)$
resonance and it substantially affects the ${\Sigma}K$
invariant mass distribution. Our results suggest that the signal 
due to the $N(1710)$ resonance excitation can be detected 
experimentally at excess energies around 100~MeV. 

Finally, we analyzed the Dalitz plot based on our  
calculations and presented the distribution of the
${\Sigma}N$ invariant mass versus the ${\Sigma}K$ invariant mass.
This analysis shows a characteristic structure  
due to the $N(1710)$ resonance that can be well distinguished from the 
lower region in the ${\Sigma}N$ invariant mass, where the final state
interaction between the $\Sigma$ and nucleon 
might be strong. However, a much cleaner signal 
of the $N(1710)$ resonance excitation can be detected 
at excess energies larger than 100~MeV, e.g. at 200~MeV.


\newpage

\end{document}